\documentclass{raa}    
\usepackage{graphicx,times}
\usepackage{misccorr}
\usepackage{booktabs} 
\usepackage{multirow}
\usepackage{natbib}
\usepackage{amssymb,amsmath}
\usepackage[nottoc,numbib]{tocbibind}
\newcommand{\specialcell}[2][c]{%
	\begin{tabular}[#1]{@{}c@{}}#2\end{tabular}}

\newcommand{\Hii}{H~{\sc ii}}

\newcommand{\kms}{km\,s$^{-1}$}

\usepackage[a4paper=true,pagebackref=true]{hyperref}

\begin{document}
	   \title{Study of the filamentary infrared dark cloud G192.76+00.10 in the S254-S258 OB complex
	}

	\volnopage{ {\bf 2018} Vol.\ {\bf X} No. {\bf XX}, 000--000}
	\setcounter{page}{1}

	\author{O.L. Ryabukhina\inst{1,2} \and
	 I. I. Zinchenko\inst{1,2} \and
	 M.R. Samal\inst{3} \and
	 P.M. Zemlyanukha\inst{1} \and
	 D.A. Ladeyschikov\inst{4} \and
	 A.M. Sobolev\inst{4} \and
	 C. Henkel\inst{5,6} \and
	 D.K. Ojha\inst{7}
	}
	
	\institute{Institute of Applied Physics of the Russian Academy of Sciences, Nizhny Novgorod, Russia; {\it ryabukhina@ipfran.ru}\\
		\and
		Lobachevsky State University of Nizhni Novgorod, Nizhny Novgorod, Russia\\
		\and
		Institute of Astronomy, National Central University, Taoyuan City, Taiwan (R.O.C.)\\
		\and
		Kourovka Astronomical Observatory, Ural Federal University, Ekaterinburg, Russia\\
		\and 
		Max Planck Institute for Radio Astronomy, Bonn, Germany\\
		\and
		Astron. Dept., King Abdulaziz University, 
  Jeddah, Saudi Arabia
		\and
		Tata Institute of Fundamental Research, Mumbai, India\\
		\vs \no
		{\small Received ...; accepted ...}
	}

\abstract{We present results of a high resolution study of the filamentary infrared dark cloud G192.76+00.10 in the S254-S258 OB complex in several molecular species tracing different physical conditions. These include three isotopologues of carbon monoxide (CO), ammonia (NH$_3$), carbon monosulfide (CS). The aim of this work is to study the general structure and kinematics of the filamentary cloud, its fragmentation and physical parameters. The gas temperature is derived from the NH$_3 $ $(J,K) = (1,1), (2,2)$ and $^{12}$CO(2--1) lines and the $^{13}$CO(1--0), $^{13}$CO(2--1) emission is used to investigate the overall gas distribution and kinematics. Several dense clumps are identified from  the CS(2--1) data. Values of the gas temperature lie in the ranges $10-35$\,K,  column density $N(\mathrm{H}_2)$ reaches the value 5.1 10$^{22}$~cm$^{-2}$. The width of the filament is of order 1~pc. The masses of the dense clumps range from $ \sim 30 $~M$_\odot$ to $ \sim 160 $~M$_\odot$. They appear to be gravitationally unstable. The molecular emission shows a gas dynamical coherence	along the filament. The velocity pattern may indicate longitudinal collapse.
\keywords{stars: formation  --- ISM: clouds --- ISM: molecules --- ISM: individual objects (G192.76+00.10) 
}
}
	
   \authorrunning{O.L. Ryabukhina et al. }            
\titlerunning{Study of the filamentary infrared dark cloud G192.76+00.10 in the S254-S258 OB complex}  
\maketitle

\section{Introduction}
	
One of the most important and actively developing areas in astrophysics is the study of star forming regions -- interstellar molecular clouds. Recent studies have shown that these clouds have a filamentary structure \citep{Andre2014}. The formation of filaments can be a necessary stage in the evolution of molecular clouds on the way to the formation of stars \citep{Andre2016}. Theoretical calculations \citep[e.g][]{Inutsuka1997} predict formation of filamentary molecular clouds after multiple compressions of interstellar gas by supersonic waves. So, areas that contain \Hii\ regions could be appropriate places for the formation of molecular filaments. An analysis of emission in different spectral lines makes it possible to comprehensively investigate the places of active star formation, as well as to evaluate their physical parameters.

\begin{figure}
	\center{\includegraphics[width=1\linewidth]{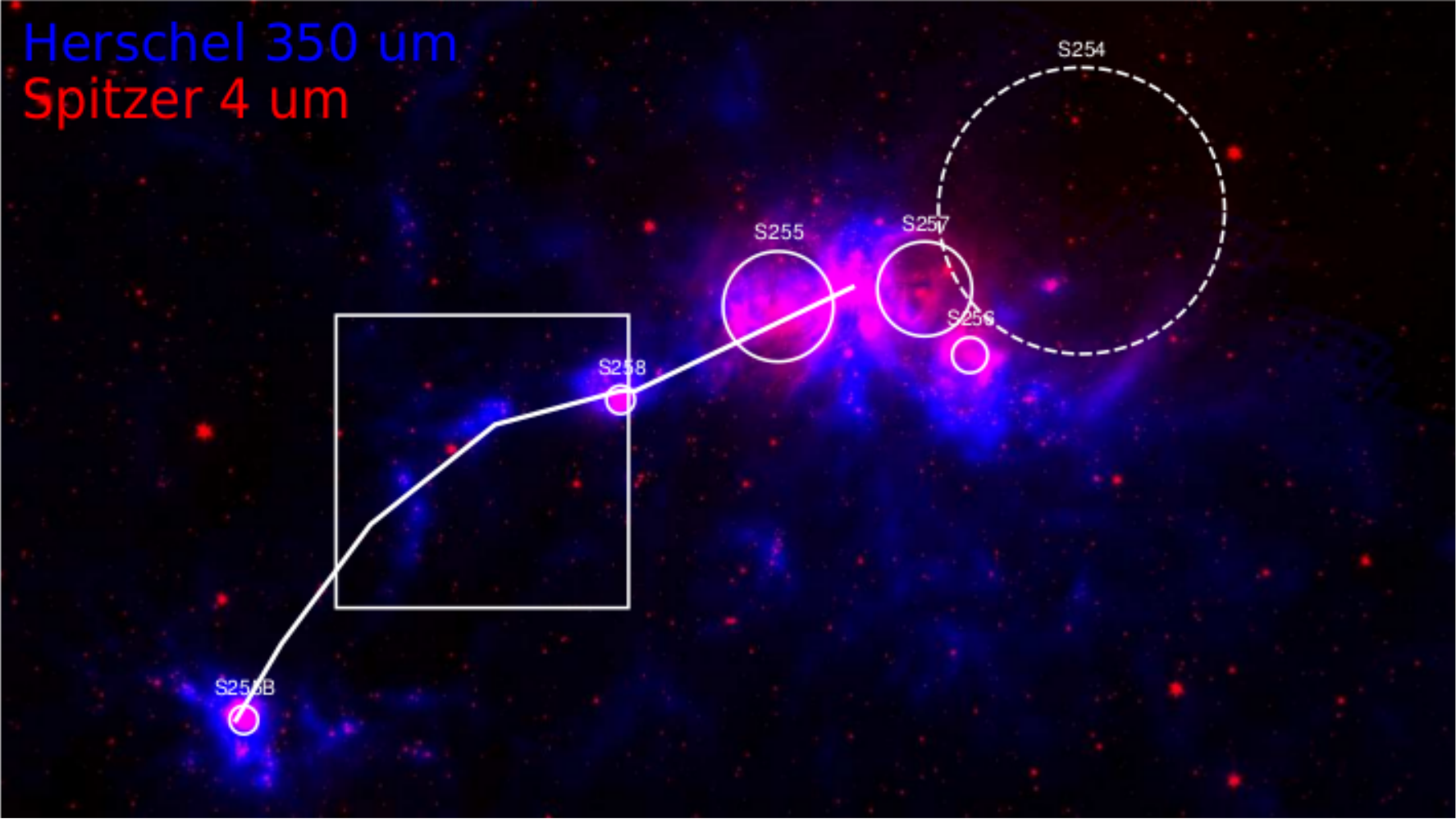}}
\caption{The map of the region S254-S258 in the infrared range at $\lambda = 350$~$\mu$m (Herschel) and $\lambda = 4$~$\mu$m (Spitzer). The white line shows the path along which the PV diagrams were constructed, white circles -- H\,II regions, the rectangular box represents the area studied in this work.} 
\label{pv_line}
\end{figure}

We study a filamentary infrared dark cloud G192.76+00.10, which is located in the star forming complex S254--S258 at a distance of $D = 1.78^{+0.12}_{-0.11}$~kpc \citep{Burns2016}. The general view of this complex in the infrared range is shown in Fig.~\ref{pv_line}. Star formation activity in this complex was investigated by \cite{Bieging2009,Chavarria2008,Ojha2011}. In the central part it may be induced by the expanding \Hii\ regions S254--S258. The cloud G192.76+00.10 was investigated by \cite{Samal15} who found that it harbours 62 YSOs distributed along the filament and the region is possibly younger than 1~Myr. They discussed that gravoturbulent fragmentation  \citep{Klessen2004} is probably the dominant cause of YSOs formation in this dark cloud.

The aim of our work is to investigate further the kinematics of this cloud, its fragmentation and physical properties at a sufficiently high resolution. For this purpose we observed this area in several molecular lines, including tracers of low and high density gas, at an angular resolution reaching $ 12^{\prime\prime} $ ($ \sim 0.1 $~pc). Here we present the observations and an analysis of the
data, including determinations of column densities, masses,
kinetic temperatures and an evaluation of the velocity
field.
	
\section{Observations} \label{obs}

In our analysis, five lines of the CO molecule and its isotopes are used, namely the  $J=1-0$ and $J=2-1$ transitions of $^{13}$CO and C$^{18}$O, and the $J=2-1$ $^{12}$CO transition. In addition, we obtained data on the $(J,K) = (1,1)$ and (2,2) NH$_3$, $J=2-1$ CS and C$^{34}$S transitions.  This set of lines makes it possible to effectively study the morphology, kinematics and physical characteristics of the molecular gas. The data reduction was performed with the XS package developed by Per Bergman at the Onsala Space Observatory and by the GILDAS software\footnote{http://www.iram.fr/IRAMFR/GILDAS}.

The data in the $^{12}$CO(2--1),  $^{13}$CO(2--1) and C$^{18}$O(2--1) lines were obtained at the 30-meter IRAM (Institut de Radioastronomie Millimetrique) radio telescope in September 2016. The observations were made with the multi-beam HERA receiver in the On-The-Fly mode and the maps are constructed with a grid spacing of 6$^{\prime \prime}$.  
Some of the HERA beams failed, which resulted in ``meshes'' in some parts of the maps (see Sects. \ref{sec:temp} and \ref{colden}).

The $^{13}$CO(1--0), C$^{18}$O(1--0), CS(2--1) and C$^{34}$S(2--1) lines were observed with the OSO (Onsala Space Observatory) 20-meter telescope  in May 2015. The NH$_3$ $(J,K) = (1,1)$ and (2,2) data were obtained with the Effelsberg telescope in April 2015. The main parameters of observations are shown in Table \ref{obs1}.

\begin{table}
	\caption{Observation parameters}
	\begin{tabular}{cccccccc}
		\toprule
		 Telescope  & Line & \specialcell{Frequency \\ (GHz)} & \specialcell{$\alpha_{2000}$ \\ (h m s)} & \specialcell{$\delta_{2000}$,\\ ($\circ $ $ \prime$ $\prime\prime$ )}& Map size & $\Theta_{FWHM}$ & \specialcell{Channel width \\ (kHz)}  \\ \midrule
		 
		\multirow{3}{*}{IRAM 30m} & $^{12}$CO (2--1) & 230.5380 & \multirow{3}{*}{6:13:40} & \multirow{3}{*}{+17:54:40} & $17^\prime \times 12^\prime$ & \multirow{3}{*}{12$^{\prime \prime}$} & \multirow{2}{*}{50} \\
		
		& $^{13}$CO (2--1) & 220.3987 & & & \multirow{2}{*}{$16^\prime \times 15^\prime$} & \\
		
		& C$^{18}$O (2--1) & 219.5604 & & & & & 200	\\ \midrule
		
		\multirow{4}{*}{ONSALA} & $^{13}$CO (1--0) & 110.2014 & \multirow{4}{*}{6:13:45} & \multirow{4}{*}{+17:54:20} & \multirow{2}{*}{$8^\prime \times 9^\prime$} & \multirow{4}{*}{36$^{\prime \prime}$} & \multirow{4}{*}{76} \\
		
		& C$^{18}$O (1--0) & 109.7822 & & &  & \\ 
		
		& CS (2--1) & 97.9810 & & & \multirow{2}{*}{$7^\prime \times 9^\prime$} &  & \\

		& C$^{34}$S (2--1) & 96.4129 & & &  &  & \\  \midrule
		
		\multirow{2}{*}{Effelsberg} & NH$_3$ (1,1) & 23.69 & \multirow{2}{*}{6:13:52} & \multirow{2}{*}{+17:53:45} & \multirow{2}{*}{$11^\prime \times 6^\prime$} & \multirow{2}{*}{33$^{\prime \prime}$} & \multirow{2}{*}{15} \\
		
		& NH$_3$ (2,2) & 23.72 & &&&& \\
		\bottomrule
	\end{tabular}
	\label{obs1}
\end{table}

\section{Results}

\subsection{The general structure of the filamentary region}

\begin{figure}
	\center{\includegraphics[width=1\linewidth]{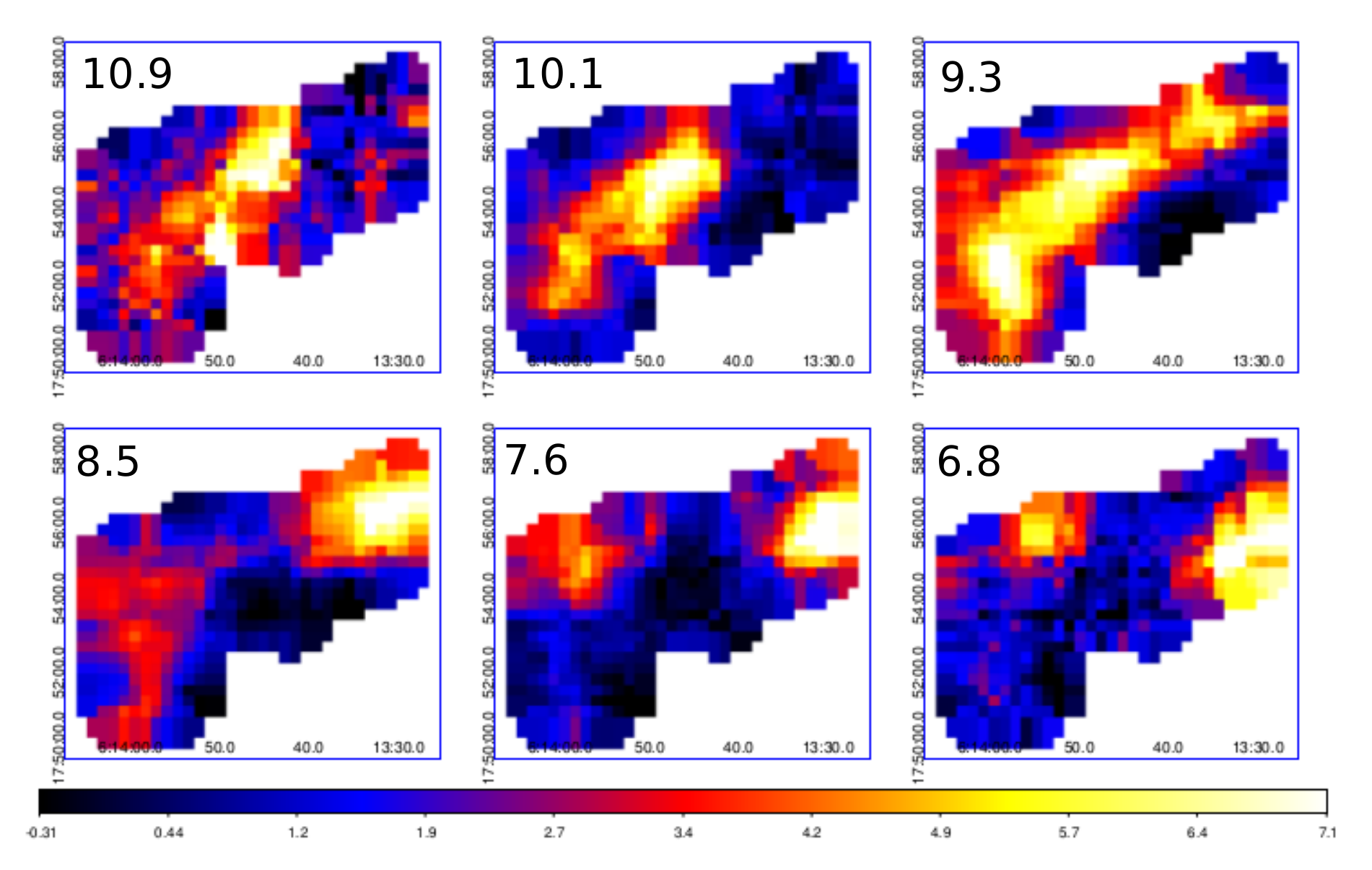}}
	\caption{Images of the G192.76+00.10 region at several velocities in the $^{13}$CO(1--0) line (ONSALA data). The color bar shows  the brightness temperature (K).The channel velocity in \kms\ is indicated in the upper left corner. }
\label{13co}
\end{figure}

The general distribution of matter at selected velocities in the $^{13}$CO(1--0) line is shown in Fig.~\ref{13co}. The channel velocity in \kms is indicated in the upper left corner. The figure shows significant velocity gradients in the area. The filamentary structure of the investigated cloud is best seen at the velocity of 9.3 km/s. At higher velocities, the middle part of the filament is observed, other parts of the cloud observed while at lower velocities. A larger scale distribution of matter as shown in the paper by  \cite{Bieging2009}.

\subsection{Kinematics}

\begin{figure}
	\center{\includegraphics[width=1\linewidth]{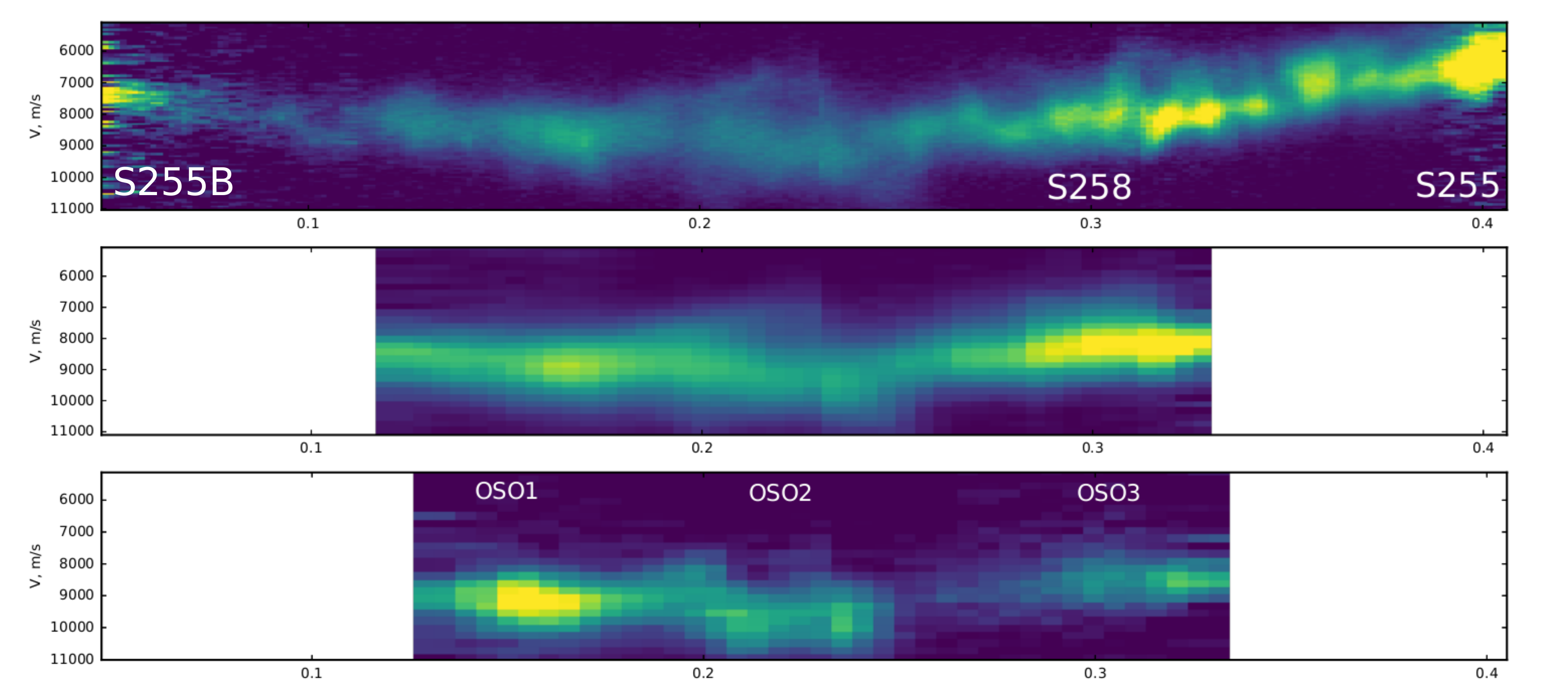}}
\caption{PV diagrams in the lines $^{13}$CO(2--1) (top), $^{13}$CO(1--0) (middle)  and CS (bottom). The horizontal axis is the distance (in degrees) along the path shown in Fig.~\ref{pv_line}.}
\label{PV}
\end{figure}

\begin{figure}
\begin{minipage}{0.5\linewidth}
	\center{\includegraphics[width=1.4\linewidth]{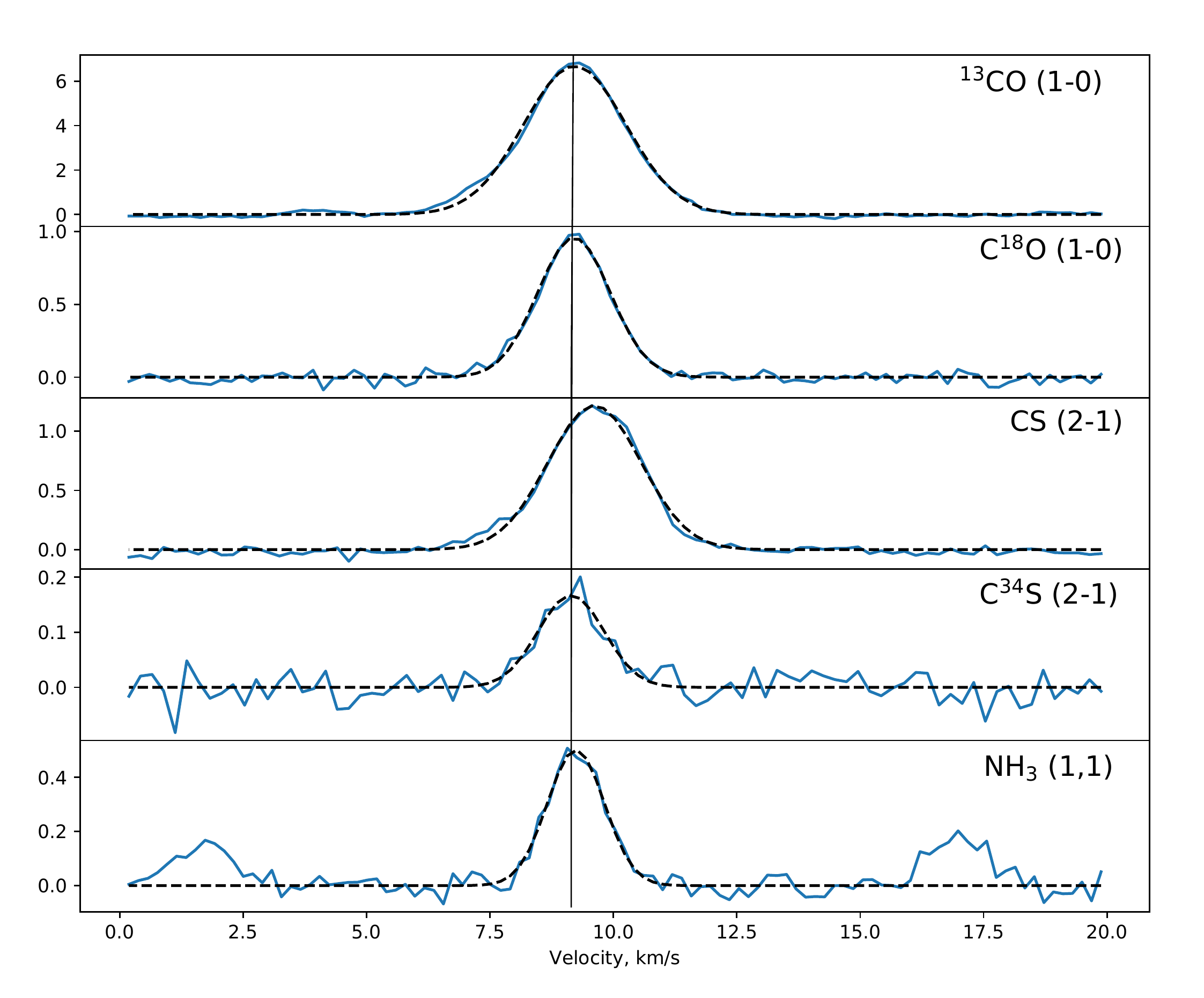}}
\end{minipage}	
\hfill
\begin{minipage}{0.3\linewidth}
	\caption{Average spectra of the $^{13}$CO (1--0), C$^{18}$O (1--0), CS (2--1), C$^{34}$S (2--1) and NH$_3$ (1,1) lines in the central part of the mapped area. The dashed lines show fitted Gaussians. The vertical line shows the velocity of the $^{13}$CO emission peak. The intensity scale is the antenna temperature (K). }
\label{spec}
\end{minipage}
	
\end{figure}

For a more detailed study of the kinematic structure of the filamentary regions, we constructed position-velocity diagrams (PV-diagrams) along the path indicated in Fig.~\ref{pv_line}, in the lines $^{12}$CO, $^{13}$CO, C$^{18}$O and CS. Position-Velocity Slice Extractor  \footnote{https://github.com/radio-astro-tools/pvextractor} was used to obtain the PV diagrams. The results for the $^{13}$CO(1--0), $^{13}$CO(2--1) and  CS(2--1) lines are shown in Fig.~\ref{PV}. We see a coherence of the line emission along the path, which confirms that this is a single entity. A gradual velocity change along the path is clearly seen. In the central part a second velocity component at lower velocities is seen, especially in the $^{13}$CO lines. The inspection of the $^{13}$CO spectra show that they are double-peaked here indeed. This part is red-shifted with respect to the ``ends" of the mapped filament. Several clumps can be distinguished, which are discussed in Sect.~\ref{clump}. The CS emission in the central part is slightly red-shifted with respect to $^{13}$CO and C$^{18}$O, however no such shift is observed in C$^{34}$S and NH$_3$. The line widths of the molecular lines are $ \sim 2 $~\kms, which greatly exceeds thermal values and implies significant turbulence in the region. Spectra of $^{13}$CO(1--0), C$^{18}$O(1--0), CS(2--1), C$^{34}$S(2--1) and NH$_3$(1,1) emission toward selected positions are shown in Fig. \ref{spec}.

\subsection{Temperature} \label{sec:temp}
The temperature is determined by two methods -- from the ammonia emission, and from the emission of the optically thick $^{12}$CO(2--1) line under the assumption of LTE conditions.

The method for evaluating the kinetic temperature from the ammonia emission in the (2,2) and (1,1) transitions is described in detail by \cite{Magnum1992}. However the ammonia emission is sufficiently strong only toward a few emission peaks.
The derived temperature is in the range of 10--20~K, increasing toward the S258 \Hii\ region.

To determine the gas temperature from the data in the $^{12}$CO (2--1) line, the method presented by \cite{Roman2010} was used. The temperature distribution map obtained by this method is shown in Fig.~\ref{temp}. The values lie within the range of 10--35 K, the highest temperatures are observed toward the S258 region. In the regions where a comparison is possible, the temperature values obtained by different methods are close to each other.

\subsection{H$_2$ column density}\label{colden}
To estimate the H$_2$ column density, we used the emission in the lines of the  $^{13}$CO (2--1) isotope, having a smaller optical depth compared to $^{12}$CO, as well as a better spatial resolution in comparison with $^{13}$CO (1--0). A number of constants was used: the CO/H$_2$ abundances ratio was taken as 8 $\times 10^{-5}$, according to   \cite{Simon2001}. The investigated region is at a distance of $D = 1.78^{+0.12}_{-0.11}$~kpc \citep{Burns2016}, which gives the galactocentric radius of 9.7~kpc or  1.21D$\odot$, if we use the distance from the Sun to center of the Galaxy  D$\odot$ = 8.34 kpc  from \cite{Reid2014}. According to \cite{Milam2005}, the $^{12}$CO/$^{13}$CO  abundances ratio at this distance is about 68, so the ratio of the abundances  $^{13}$CO/H$_2$ = [CO/H$_2$]/[$^{12}$CO/$^{13}$CO] $\sim$ 1.17 $\times 10^{-6}$. This value was used in evaluating the H$_2$ column density and the masses of clumps (see Sects. \ref{clumps}).

\begin{figure}
\begin{minipage}{0.49\linewidth}
	\center{\includegraphics[width=1\linewidth]{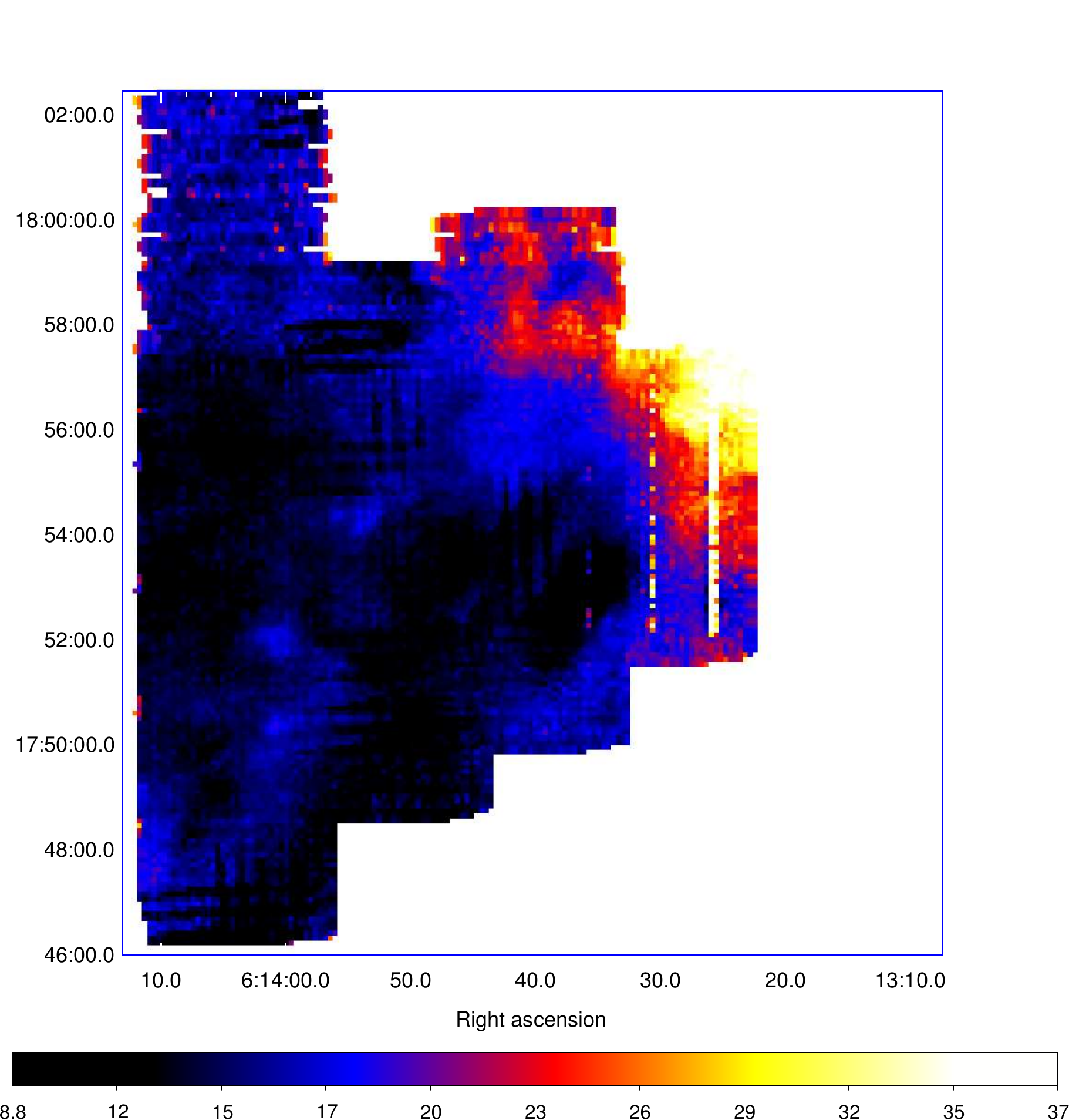}}
		\caption{The temperature distribution map, derived from the $^{12}$CO (2-1) emission (IRAM data). The ``meshes" on the map are the instrumental effects caused by the fact that some of the beams of the receiver (HERA) failed.}
		\label{temp}
\end{minipage}	
\hfill
\begin{minipage}{0.49\linewidth}
	\center{\includegraphics[width=1\linewidth]{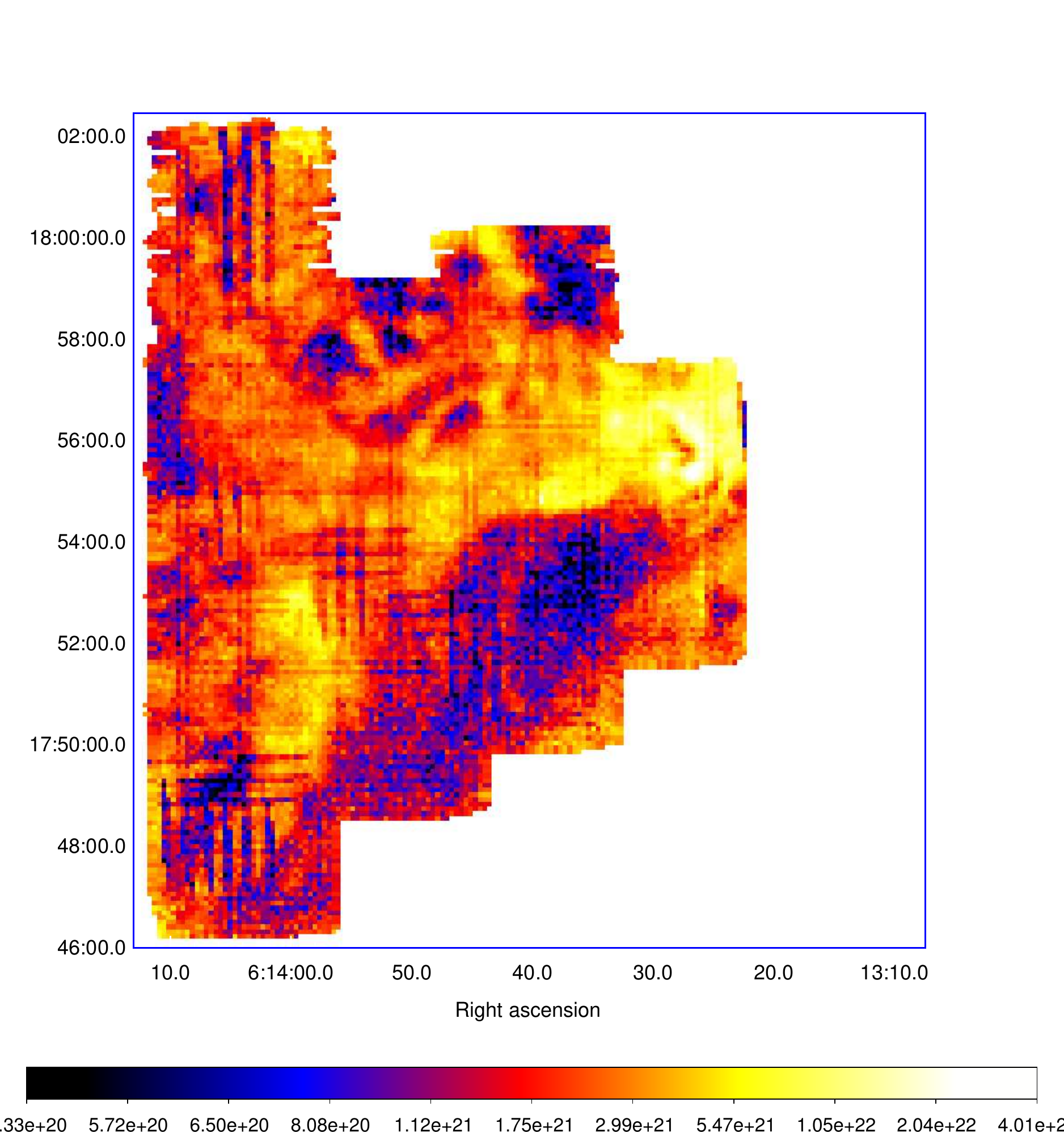}}
	\caption{The H$_2$ column density map, derived from the $^{13}$CO(2-1) emission (IRAM data). The ``meshes" on the map are the instrumental effects caused by the fact that some of the beams of the receiver (HERA) failed.}
	\label{NH2}
\end{minipage}	
\end{figure}

Next, we estimate the optical depth in the $^{13}$CO (2--1) line by the formula (15.31) from  \cite{Rohlfs2004}:

\begin{equation}
\tau^{13}_0 = -ln \left[1 - \frac{T^{13}_B / T_0}{(e^{T_0/T_{ex}}-1)^{-1}-(e^{T_0/2.7}-1)^{-1}} \right]
\label{Tau}
\end{equation}
where $T^{13}_B$ is the brightness temperature in the $^{13}$CO line and $T_{ex}$ is the excitation temperature obtained from $^{12}$CO (Sect.~\ref{sec:temp}). Assuming Local Thermodynamical Equilibrium
(LTE) and accounting for the fact that $^{13}$CO is a linear
molecule, the column density is related (equation (15.37) from \cite{Rohlfs2004})

\begin{equation}
N(^{13}\mathrm{CO}) = 1.5 \times 10^{14} \frac{e^{5.3/T_{ex}}}{1-e^{-10.6/T_{ex}}} \times T_{ex} \int \tau^{13}(v) dv
\label{N13CO}
\end{equation}

for the transition of $^{13}$CO(2--1). Further, using the known $^{13}$CO abundance, we obtain the $N_{\mathrm{H}_2}$ column density. The $N(\mathrm{H}_2)$ distribution is presented in Fig.~\ref{NH2}. The column density is in the range from 6.2\,10$^{20}$ to 5.1\,10$^{22}$~cm$^{-2}$. 

	\subsection{Filament width}\label{width}

\begin{figure}
\begin{minipage}{0.6\linewidth}
	\center{\includegraphics[width=1.23\linewidth]{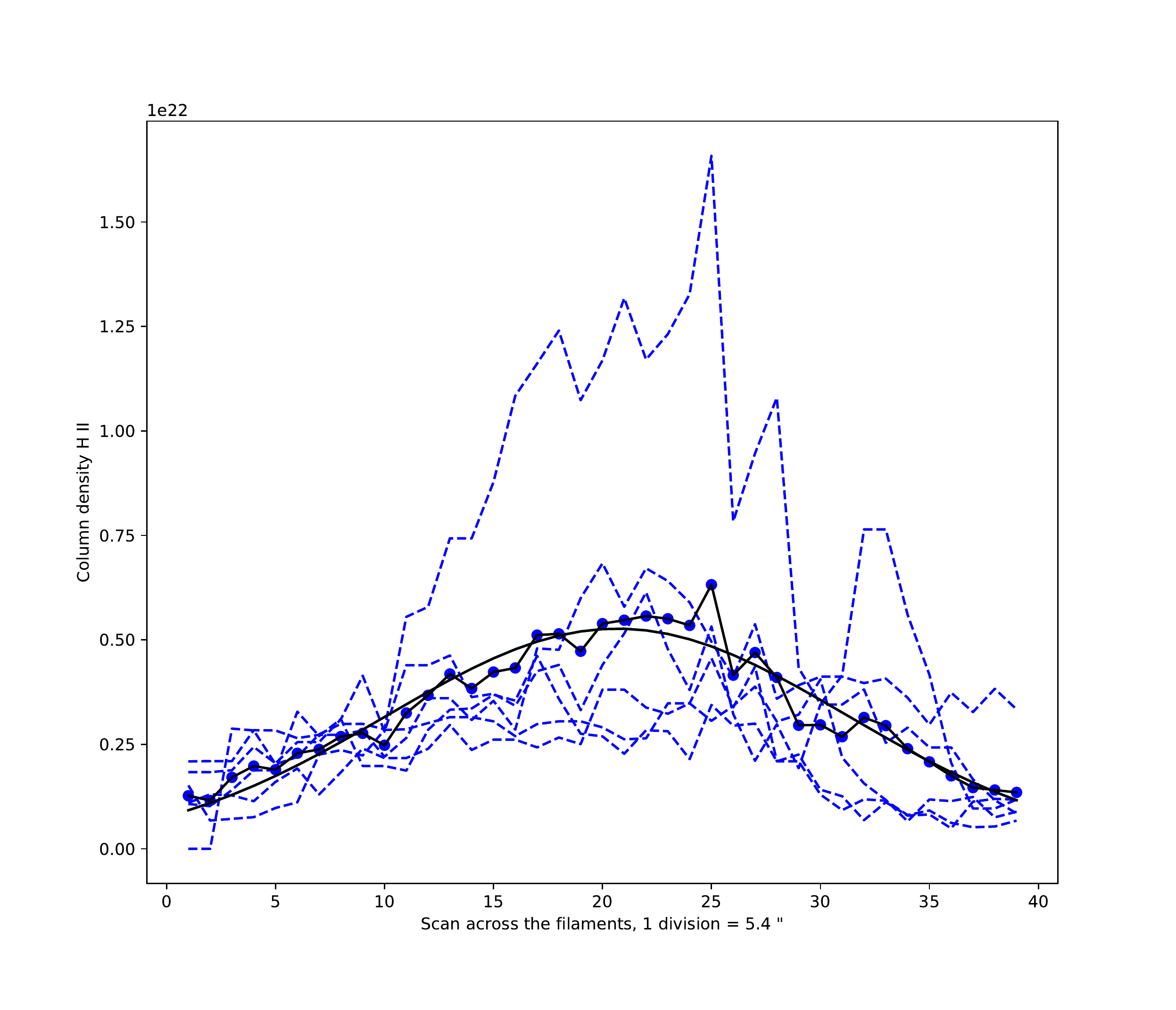}}
\end{minipage}	
\hfill
\begin{minipage}{0.3\linewidth}
	\caption{The profiles of the H$_2$ column density for 6 cuts perpendicular to the filament (dashed lines), the averaged profile (solid line marked by circles) and fitted Gaussian (solid line).}
	\label{widgh}
\end{minipage}
\end{figure}

To determine the width of the filament, we used the distribution of the H$_2$ column density (Sect.~\ref{colden}). Profiles of the column density along 6 lines perpendicular to the filament were constructed  (Fig.~\ref{widgh}),  these data were averaged, and using the GaussianModel algorithm of the LMFIT module, \footnote{https://lmfit.github.io/lmfit-py/} the Gaussian function is fitted. The deconvolved full width at the half maximum level is $0.98 \pm 0.03$~pc. It is worth mentioning that the widths for different cuts are rather similar.

\subsection{Filament mass}\label{mass}

Knowing the distribution of the column density H$_2$ (Section~\ref{colden}), we obtain the mass of the gas by integrating the $N(\mathrm{H}_2)$column density over the source surface:

\begin{equation}
M = \mu m_{H_2} \int N_{H_2} dA = \mu m_{H_2} D^2 \int N_{H_2} d\Omega
\label{mass}
\end{equation}

where $ \mu $ is the average molecular weight with respect to the mass of the hydrogen molecule \citep{Kauffmann2008}, and the surface element $dA$ is connected with the solid angle by the relation $dA =  D^2d\Omega $, where $D$ is the distance to the source. 

According to these calculations, the mass of the investigated filament region is  $\sim 800$~M$_\odot$, and the length is $\sim 7$~pc. Mass per unit length comes out to be $\sim$ 115 M$_\odot$/pc, which exceeds $M_{crit} = 2 c_s^2/G \sim 25$~M$_\odot$/pc \citep{Samal15}, where c$_s$ is the sound speed of the medium, and $G$ is the gravitational constant. 

\subsection{Identification of dense clumps}\label{clumps}

To identify dense molecular clumps, we use the GaussClumps algorithm, first proposed by \cite{Stutzki1990}. In the data cube of the Position-Position-Velocity (PPV) type, the absolute maximum of the emission is determined, after which a three-dimensional Gaussian is fitted into the position of this maximum, which is then subtracted from the original cube. After that, the next maximum is searched, followed by fitting and subtraction. This procedure continues until the criterion for the completion of the algorithm is satisfied. 

\begin{figure}
\begin{minipage}{0.49\linewidth}
	\center{\includegraphics[width=1\linewidth]{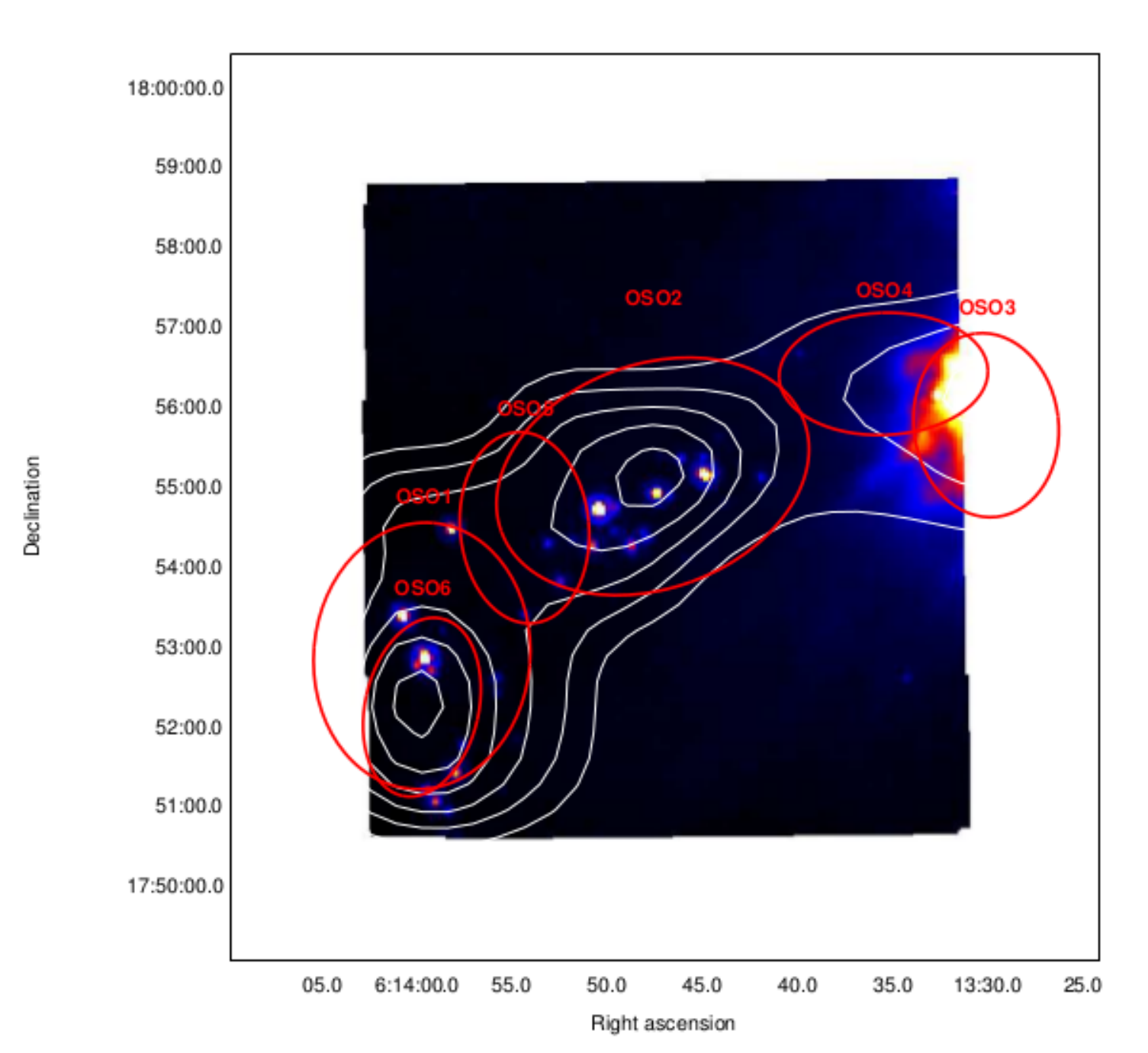}}
	\caption{Image shows Spitzer MIPS 24~$\mu$m emission. White contours represent integrated CS(2--1) emission and red ellipses show clumps identified with the GaussClump procedure (Table \ref{onsala}).}
	\label{spitzer}
\end{minipage}
\hfill
\begin{minipage}{0.49\linewidth}
	\center{\includegraphics[width=1\linewidth]{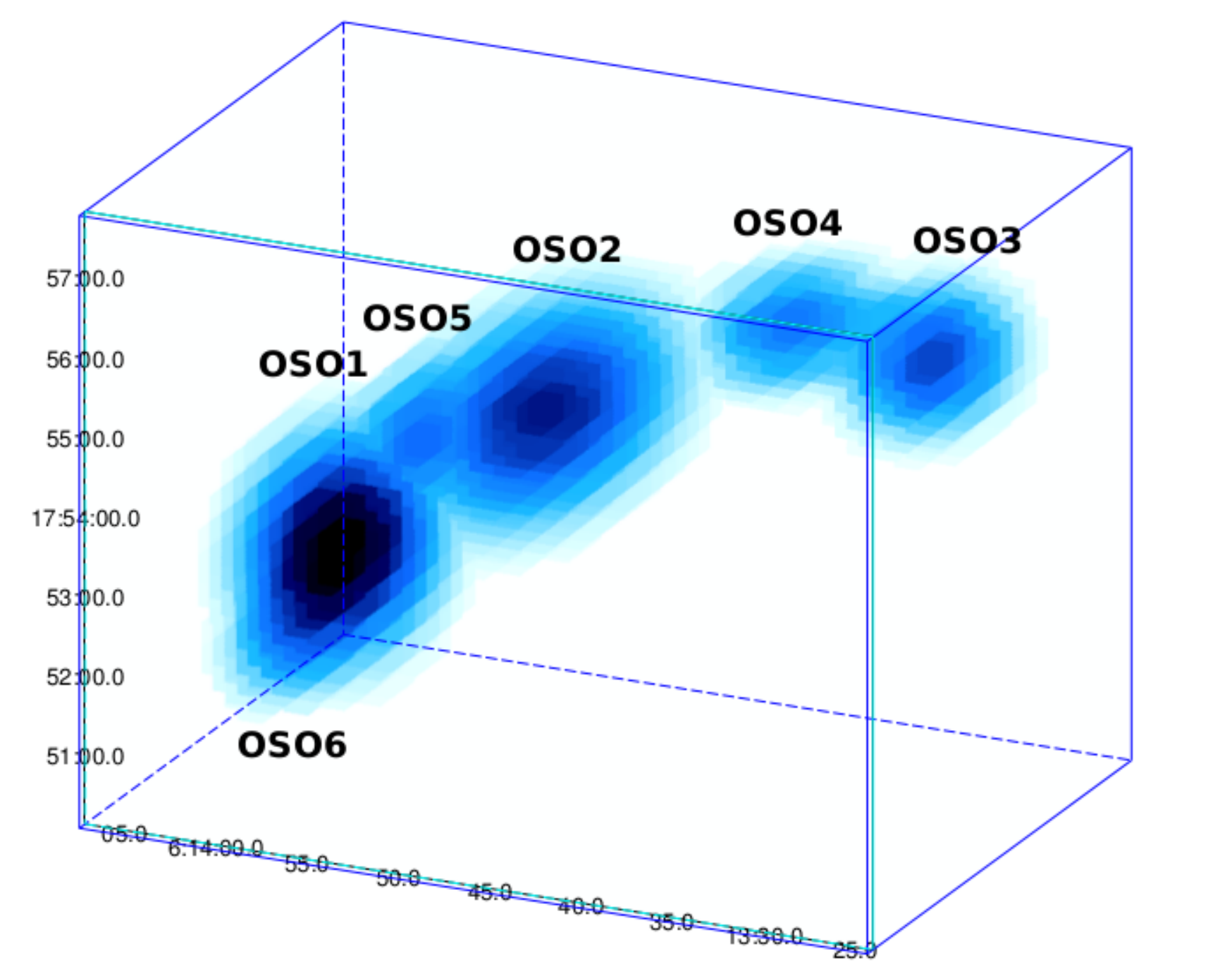}}
	\caption{3D visualization of the clumps identified with GaussClump.}
	\label{clump}
\end{minipage}
\end{figure}

The CS (2--1) emission was used to identify the clumps as a traditional dense gas tracer.  Six clumps were found and the following parameters of the algorithm completion were used: FWHM of the instrument beam in pixels (FwhmBeam) = 1.5, FWHM in velocity -- 0.7~\kms. The dimensions of the clumps are defined as the widths at the half-intensity level $ \Theta_{FWHM} $. Visualization of clumps is shown in Fig.~\ref{spitzer}, 3D visualization as shown in Fig.~\ref{clump}. Clumps obtained by CS coincide with star clusters.

\subsection{Physical parameters of clumps}
To derive the masses of the clumps, the method presented in Section~\ref{mass} was used, however, the integration in Eq.~\ref{N13CO} was performed only at velocities at which the clumps emission is observed. 

The virial parameter of the clumps $\alpha_{vir} = M_{vir}/M$ is calculated according to the definition in \cite{Kauffmann2013}:

\begin{equation}
\alpha_{vir} = \frac{5 \sigma_v^2 R}{G M} = 1.2 \left(\frac{\sigma_v}{km s^{-1}}\right)^2 \left(\frac{R}{pc}\right) \left(\frac{M}{M_\odot}\right)
\label{vir}
\end{equation}

 where $\sigma_v$ is the velocity dispersion, $R$ is the clump radius and $G$ is the gravitational constant.
 
 The parameters of the clumps are indicated in Table  \ref{onsala}, where  $\alpha$, $\beta$, V$_{max}$ are the coordinates of the clumps, $\Theta_{FWHM}$ is the width of the fitted Gaussian in angular minutes,  max $N(\mathrm{H}_2)$ is the maximum value of the hydrogen column density, $M$/M$_\odot$ and $M_{vir}$/M$_\odot$ are the mass and virial mass in units of solar masses and $\alpha_{vir}$ is the virial parameter.

\begin{table}
	\caption{Clumps identified in the CS(2-1) line.}
	\begin{tabular}{ccccccccc}
		\toprule
		Clump  & \specialcell{$\alpha_{2000}$,\\ (h m s)} & \specialcell{$\delta_{2000}$,\\ ($\circ $ $ \prime$ $\prime\prime$) }& \specialcell{V$_{peak}$ \\ (km/s)} & \specialcell{$\Theta_{FWHM}$ \\ (arcmin)} & \specialcell{max N$_{H_2}$\\ (10$^{21}$ cm$^{-2}$)} & \specialcell{ M \\ (M$_\odot$)} & \specialcell {M$_{vir}$ \\ (M$_\odot$)} & $\alpha_{vir}$ \\ \midrule
		OSO 1 & 6:13:59.7 & +17:52:50.0 & 9.328 &2.7$\times$3.38 & 12.8 & 66 & 48 & 0.72 \\
		OSO 2 & 6:13:47.5 & +17:55:04.5 & 9.562 & 4.05$\times$2.7 & 7.27 &  161 & 125 & 0.77 \\
		OSO 3 & 6:13:30.0 & +17:55:43.1 & 8.161 & 2.02$\times$2.37 & 51.3 & 162 & 28 & 0.17 \\
		OSO 4 & 6:13:35.4 & +17:56:20.9 & 9.095 & 2.7$\times$1.6 & 13.5 & 88 & 14 & 0.16 \\
		OSO 5 & 6:13:54.3 & +17:54:26.0 & 9.562 & 1.69$\times$2.4 & 3.56 & 32 & 25 & 0.78 \\ 
		OSO 6 & 6:13:59.7 & +17:52:10.9 & 10.26 & 1.6$\times$2.4 & 5.48 &  30 & 13 & 0.43 	\\\bottomrule
	\end{tabular}
\label{onsala}
\end{table}

\section{Discussion}

\cite{Chavarria2008} have shown that clusters of young stellar objects of the star formation complex S254--S258 are located at the boundaries of the \Hii\ regions. Based on the molecular gas distribution analysis, \cite{Bieging2009} also conclude that the star formation processes in this region  is related to  the expansion of neighboring \Hii\ regions. These processes are reflected in the large-scale structure and kinematics of the star-forming regions. 

The general morphology of the investigated region, as seen in the H$_2$ column density map (Fig.~\ref{NH2}), is rather complicated. In addition to the ``main" filament discussed here, there is a filamentary structure of lower column density in the north-eastern part, which intersects with the main one and their interaction is possible.

The velocity pattern seen along the main filament allows different interpretations. Similar velocity structure was seen in some other cases \citep[e.g.][]{Peretto2014, Hakar2017, Kirsanova2017} and was interpreted as an evidence for a filament's longitudinal collapse. Such a possibility looks probable here, too.

The filaments's width ($\sim$ 1\,pc)  obtained in Section~\ref{width} is significantly larger than the average values for interstellar filaments of various types \citep[e.g.][]{Andre2016, Arzoumanian2011, Li2016} but is not exceptional. Theoretical models \citep[e.g.][]{Hartmann02} show that the radial scale height of a filament, in case of thermal pressure support, is determined by the sound speed and the surface density. An additional turbulent pressure support may increase this scale. Our line widths are certainly non-thermal, so the variant of a turbulent support, as a reason of the large filament width, seems to be probable.

A filamentary cloud is unstable if its mass ratio per unit length is greater than the critical ratio  $M_{line} > M_{crit} = 2 c_s^2/G$, where $c_s$ is the sound speed, $G$ is the gravitational constant \citep{Inutsuka1997}. For the investigated region we find $M_{line} \sim 115$~M$_\odot$ and $M_{crit} \sim 25$~M$_\odot$/pc \citep{Samal15}. 

To determine the column density and mass of the clumps, we followed the procedure presented in \cite{Rohlfs2004}. The parameters of the selected clumps are presented in the Table~\ref{onsala}. Masses of clumps lie in the range of 30--160 M$_\odot$, the value of the virial parameter varies from 0.16 in the clump OSO4 to 0.78 in OSO5. \cite{Kauffmann2013} have shown that if the virial parameter $\alpha_{vir} > \alpha_{crit}$, then the clump or molecular cloud is gravitationally stable. If $\alpha_{vir} \lesssim \alpha_{crit}$, then the perturbations of the pressure and density of the clumps can lead to a gravitational contraction and start of the processes of star formation. For isothermal clumps without taking into account the presence of magnetic fields $\alpha_{crit} \simeq$ 2 \citep{Kauffmann2013}. For all the studied clumps, the virial parameter satisfies this condition, which implies  their gravitational instability. The OSO3 and OSO4 clumps have the smallest virial parameters, than in other clumps, which is due to the small dimensions of these clumps and the high H$_2$ column density. 
 
\section{Conclusions}
The main results of this study are the following:

1. It is shown that the filamentary dark cloud in the G192.76+00.10 region is dynamically coherent. The shape of the position-velocity diagrams may imply the gas accretion along the filament to its central part.

2. The gas temperature determined from the $ ^{12}$CO and NH$_3$ emission is $ 10-35 $~K. 

3. The hydrogen column density reaches the value 5.1 10$^{22}$~cm$^{-2}$. The total mass of the investigated part of the filament is $\sim 800$~M$_\odot$, the length is $\sim 7$~pc. The mass per unit length is $\sim$ 115 M$_\odot$/pc, which is higher than the critical value and indicates gravitational instability in the absence of a stabilizing magnetic field.

4. The average width of the filament obtained from the gas column density distribution is about 1~pc, much larger than the average values for interstellar filaments. It may be related to an additional turbulent pressure support.

5. Six dense clumps are identified in the CS(2--1) emission and their physical parameters are determined. Masses of the clumps lie within the range of $30-160$~M$_\odot $, the value of the virial parameter range from 0.16 to 0.78, which implies their gravitational instability.

\section{Acknowledgments}

This research was supported by the Russian Foundation for Basic Research (grants No. 15-02-06098 and 17-52-45020) in the part of the observations and preliminary data reduction, and by the Russian Science Foundation (grant No. 17-12-01256) in the part of the data analysis. We are grateful to the anonymous referee for the helpful comments and suggestions.
 
\bibliographystyle{raa}
\bibliography{biblio}

\begin{thebibliography}{24}
\providecommand\natexlab[1]{#1}
\providecommand\JournalTitle[1]{#1}

\bibitem[{Andr{\'e}} {et~al.}(2014)]{Andre2014}
{Andr{\'e}}, P., {Di Francesco}, J., {Ward-Thompson}, D., {et~al.} 2014,
  Protostars and Planets VI, 27

\bibitem[{Andr{\'e}} {et~al.}(2016)]{Andre2016}
{Andr{\'e}}, P., {Rev{\'e}ret}, V., {K{\"o}nyves}, V., {et~al.} 2016, \aap,
  592, A54

\bibitem[{Arzoumanian} {et~al.}(2011)]{Arzoumanian2011}
{Arzoumanian}, D., {Andr{\'e}}, P., {Didelon}, P., {et~al.} 2011, \aap, 529, L6

\bibitem[{Bieging} {et~al.}(2009)]{Bieging2009}
{Bieging}, J.~H., {Peters}, W.~L., {Vila Vilaro}, B., {Schlottman}, K., \&
  {Kulesa}, C. 2009, \aj, 138, 975

\bibitem[{Burns} {et~al.}(2016)]{Burns2016}
{Burns}, R.~A., {Handa}, T., {Nagayama}, T., {Sunada}, K., \& {Omodaka}, T.
  2016, \mnras, 460, 283

\bibitem[{Chavarr{\'{\i}}a} {et~al.}(2008)]{Chavarria2008}
{Chavarr{\'{\i}}a}, L.~A., {Allen}, L.~E., {Hora}, J.~L., {Brunt}, C.~M., \&
  {Fazio}, G.~G. 2008, \apj, 682, 445

\bibitem[{Hacar} {et~al.}(2017)]{Hakar2017}
{Hacar}, A., {Alves}, J., {Tafalla}, M., \& {Goicoechea}, J.~R. 2017, \aap,
  602, L2

\bibitem[{Hartmann}(2002)]{Hartmann02}
{Hartmann}, L. 2002, \apj, 578, 914

\bibitem[{Inutsuka} \& {Miyama}(1997)]{Inutsuka1997}
{Inutsuka}, S.-i., \& {Miyama}, S.~M. 1997, \apj, 480, 681

\bibitem[{Kauffmann} {et~al.}(2008)]{Kauffmann2008}
{Kauffmann}, J., {Bertoldi}, F., {Bourke}, T.~L., {Evans}, II, N.~J., \& {Lee},
  C.~W. 2008, \aap, 487, 993

\bibitem[{Kauffmann} {et~al.}(2013)]{Kauffmann2013}
{Kauffmann}, J., {Pillai}, T., \& {Goldsmith}, P.~F. 2013, \apj, 779, 185

\bibitem[{Kirsanova} {et~al.}(2017)]{Kirsanova2017}
{Kirsanova}, M.~S., {Salii}, S.~V., {Sobolev}, A.~M., {et~al.} 2017,
  arXiv:1711.01428

\bibitem[{Klessen} {et~al.}(2004)]{Klessen2004}
{Klessen}, R.~S., {Ballesteros-Paredes}, J., {Li}, Y., \& {Mac Low}, M.-M.
  2004, in Astronomical Society of the Pacific Conference Series, Vol. 322, The
  Formation and Evolution of Massive Young Star Clusters, ed. H.~J.~G.~L.~M.
  {Lamers}, L.~J. {Smith}, \& A.~{Nota}, 299

\bibitem[{Li} {et~al.}(2016)]{Li2016}
{Li}, G.-X., {Urquhart}, J.~S., {Leurini}, S., {et~al.} 2016, \aap, 591, A5

\bibitem[{Mangum} {et~al.}(1992)]{Magnum1992}
{Mangum}, J.~G., {Wootten}, A., \& {Mundy}, L.~G. 1992, \apj, 388, 467

\bibitem[{Milam} {et~al.}(2005)]{Milam2005}
{Milam}, S.~N., {Savage}, C., {Brewster}, M.~A., {Ziurys}, L.~M., \& {Wyckoff},
  S. 2005, \apj, 634, 1126

\bibitem[{Ojha} {et~al.}(2011)]{Ojha2011}
{Ojha}, D.~K., {Samal}, M.~R., {Pandey}, A.~K., {et~al.} 2011, \apj, 738, 156

\bibitem[{Peretto} {et~al.}(2014)]{Peretto2014}
{Peretto}, N., {Fuller}, G.~A., {Andr{\'e}}, P., {et~al.} 2014, \aap, 561, A83

\bibitem[{Reid} {et~al.}(2014)]{Reid2014}
{Reid}, M.~J., {Menten}, K.~M., {Brunthaler}, A., {et~al.} 2014, \apj, 783, 130

\bibitem[{Rohlfs} \& {Wilson}(2004)]{Rohlfs2004}
{Rohlfs}, K., \& {Wilson}, T.~L. 2004, {Tools of radio astronomy}

\bibitem[{Roman-Duval} {et~al.}(2010)]{Roman2010}
{Roman-Duval}, J., {Jackson}, J.~M., {Heyer}, M., {Rathborne}, J., \& {Simon},
  R. 2010, \apj, 723, 492

\bibitem[{Samal} {et~al.}(2015)]{Samal15}
{Samal}, M.~R., {Ojha}, D.~K., {Jose}, J., {et~al.} 2015, \aap, 581, A5

\bibitem[{Simon} {et~al.}(2001)]{Simon2001}
{Simon}, R., {Jackson}, J.~M., {Clemens}, D.~P., {Bania}, T.~M., \& {Heyer},
  M.~H. 2001, \apj, 551, 747

\bibitem[{Stutzki} \& {Guesten}(1990)]{Stutzki1990}
{Stutzki}, J., \& {Guesten}, R. 1990, \apj, 356, 513

\end{thebibliography}

\end{document}